\documentclass[amsmath, preprintnumbers, 10pt, twocolumn, prl,longbibliograpy]{revtex4-1}

\usepackage{graphicx}
\usepackage{subfigure}

\begin{document}
\title{Interface height fluctuations and surface tension of driven liquids with time-dependent dynamics}
\author{Clara \surname{del Junco} and Suriyanarayanan Vaikuntanathan}
\affiliation{Department of Chemistry and The James Franck Institute, University of Chicago, Chicago, IL, 60637}

\begin{abstract}
Interfaces in phase-separated driven liquids are one example of how energy input at the single-particle level changes the long-length-scale material properties of nonequilibrium systems. Here, we measure interfacial fluctuations in simulations of two liquids driven by time-dependent forces, one with repulsive interactions and one with attractive interactions. The time-dependent forces lead to currents along the interface, which can modify the scaling of interface height fluctuations with respect to predictions from capillary wave theory (CWT). We therefore characterize the whole spectrum of fluctuations to determine whether CWT applies. In the system with repulsive interactions, we find that the interface fluctuations are well-described by CWT at one amplitude of the driving forces but not at others. In the system with attractive interactions, they obey CWT for all amplitudes of driving, allowing us to extract an effective surface tension. The surface tension increases linearly over two orders of magnitude of the driving forces, more than doubling its equilibrium value.  Our results show how the interfaces of nonequilibrium liquids with time-dependent forces are modified by energy input.
\end{abstract}

\maketitle

\section{Introduction}

In recent years there have been many reports of experimental~\cite{Han2017,Palacci2013, Kumar2014} and simulated~\cite{Fily2012, Redner2013, Cates2015, Liebchen2016b, Nguyen2014b, Yeo2015b} particle systems with purely repulsive interactions that are always homogeneous at equilibrium but undergo phase separation when driven out of equilibrium. Understanding how non-equilibrium driving modifies interfacial fluctuations in these cases - and material properties in general - is an important and open question. For instance, surface fluctuations play a central role in micro-scale applications~\cite{Sackmann2014}, and understanding how to control them can contribute to our ability to exploit the engineering promise of nonequilibrium particle systems~\cite{Whitesides2002, Grunwald2016a, Grzybowski2017}.

A few examples of nonequilibrium phase separation are motility-induced phase separation (MIPS), undergone by Brownian particles when they are given the ability to self-propel~\cite{Fily2012, Redner2013, Bialke2013, Cates2015, Stenhammar2015, Whitelam2018}, lane or stripe formation of charged particles in an electric field~\cite{Dzubiella2002, Wysocki2009, Vissers2011, Vissers2011a, Klymko2016} and shaken granular matter~\cite{Mullin2000, Pooley2004}, and the separation of particles with rotational dynamics based on phase synchronization~\cite{Liebchen2017} or chirality~\cite{Nguyen2014b, Han2017, Yeo2016}.
Recently we reported phase separation of this last kind and stable, system-spanning interfaces in simulations of a liquid of 2-dimensional disks with repulsive interactions where half of the particles are driven by a time dependent field so that they orbit in phase~\cite{delJunco2018}.  This model was inspired by a recent experimental study  in which magnetic particles are driven by a rotating magnetic field and undergo phase separation~\cite{Han2017}. Here, we combine simulations with an analysis based on capillary wave theory (CWT)~\cite{Bedeaux1985}  to study the effect of the time-dependent forces on the interfacial properties of the liquid with repulsive interactions, and of a closely related liquid with attractive interactions. To distinguish these systems, we will refer to the repulsive model studied in Ref.~\citenum{delJunco2018} as the Weeks-Chandler-Andersen (WCA) model, and to the new attractive model as the Lennard-Jones (LJ) model.

The main result of CWT predicts that the power spectrum of height fluctuations of an interface parallel to a prescribed horizontal axis satisfies $\langle |h(k)|^2\rangle \propto 1/ (\sigma k^2)$, where $k$ denotes the wavevector, $h(k)$ denotes the Fourier transform of height fluctuations, and $\sigma$ is the surface tension. This $1/k^2$ scaling - also known as capillary scaling - is found in systems ranging from the 2D Ising model to water~\cite{Fisher1982, Sides1999,Schwartz1990a}. CWT has also been used to study interfaces in non-equilibrium liquids and extract effective surface tensions~\cite{Bialke2015, Paliwal2017, Derks2006, Patch2018, Lee2017}. 

It has been shown that phase separation in the WCA model belongs to the Ising universality class~\cite{Han2017}, which would lead us to expect capillary scaling of the interface modes~\cite{Fisher1982}. This expectation is brought in to question by our first finding, which is that the time-dependent driving forces result in persistent particle currents along the interface of the WCA model (Fig.~\ref{fig:velocityprof}). These can affect the statistics of interface fluctuations. For instance, the currents present in a non-equilibrium Ising model with an applied electric field can cause the scaling to decrease to $1/k^{0.67}$\cite{Leung1993}.

The fluctuations of active interfaces have been studied recently in systems of the MIPS type~\cite{Bialke2015, Paliwal2017, Lee2017, Patch2018} where there can be local tangential flows~\cite{Patch2018} but not system-spanning currents at the interface. In our system, and in others with rotational dynamics, we observe system spanning currents qualitatively similar to those in the non-equilibrium driven Ising model~\cite{Leung1993}. The presence of these currents makes it important to examine the full spectrum of capillary fluctuations. This examination will allow us to assess whether the system obeys capillary scaling and for what range of wavenumbers, to check the convergence of interface statistics, and to accurately measure the surface tension.

In the WCA model, we find that the scaling of interface fluctuations depends on the amplitude of the driving forces. For one amplitude that we studied, we find close to $1/k^2$ scaling, while for all others we find that $\langle |h(k)|^2\rangle$ is inversely correlated with $k$, but decreases less rapidly than predicted by CWT (Fig.~\ref{fig:hk_WCA}). The effect of the driving forces on the stability of interfaces in the WCA model is non-monotonic, because they cause the system to phase separate at low amplitudes but to become mixed again at large amplitudes~\cite{delJunco2018}. Moreover, since the system is mixed at equilibrium, there is no reference value for the surface tension in the absence of driving. For these reasons, the WCA model is not ideal to systematically investigate the effect of driving forces on surface tension.

For this purpose, we introduce the LJ model, which is phase-separated with a well-defined surface tension at equilibrium~\cite{Paliwal2017}. We find that the LJ model exhibits capillary fluctuations over a wide range of wavevectors $k$ even in the presence of driving. Over an order of magnitude in the driving forces, the effect of driving in the LJ model is a linear increase in the surface tension ($\sigma$).  We discuss two ways that the driving forces can increase the force imbalance at the interface, thereby causing the observed increase in $\sigma$: first, by inducing a restoring force on the interface that is proportional to the curvature, and second, by changing the density of the liquid and gas phases of LJ particles. We show that both of these effects can contribute to the increase in the surface tension, but a full account of the linear trend remains an open problem that these equilibrium-like arguments are insufficient to address.
% Need a final sentence

\section{Methods}

\subsection{Models and Simulation Details}

We studied interfaces in two models of driven liquids: one in which the particles have repulsive interactions only, which does not phase separate at equilibrium, and a second with attractive interactions between driven particles and repulsive interactions between undriven particles, which phase separates and possesses stable interfaces at equilibrium.  
Both models consist of 2-dimensional disks whose positions evolve in time according to driven Brownian dynamics:
\begin{equation}
{\bf \dot r}_i(t) = D_0\beta\left({\bf F}_{ c,i}(t) + {\bf F}_{ d}(t)\right)+\boldsymbol\eta_i (t).
\label{eq:xEOM}
\end{equation}
Here, $D_0$ is the diffusion constant of a single particle and $\boldsymbol\eta_i (t) = (\eta_{i,x}(t),\eta_{i,y}(t))$ are Gaussian-distributed random variables with $\left<\boldsymbol\eta_i (t)\right>=0$ and $\left<\eta_{i,\mu}(t)\eta_{j,\nu}(t')\right> = 2D_0\delta_{i,j}\delta_{\mu,\nu}\delta(t-t')$. $D_0$ is related to the friction coefficient $\gamma$ by $D_0 = k_BT/\gamma$. In all of our simulations and calculations, we set $\beta=(k_BT)^{-1}=1$. The length scale of the system is set by the particle diameter, $r_0$, and the time scale is set by $t_0 = D_0/r_0^2$.

In the WCA model, previously described in Ref.~\citenum{delJunco2018} and motivated by Ref.~\citenum{Han2017}, ${\bf F}_{ c,i}$ is the (purely repulsive) conservative force on particle $i$ due to the Weeks-Chandler-Andersen interaction potential~\cite{WCA1971}:
\begin{equation} 
u(r_{ij})=\left\{ \begin{matrix} 4\epsilon_{WCA} \left[ \left( \frac{r_0}{r_{ij}} \right)^{12}-\left( \frac{r_0}{r_{ij}}\right)^{6}\right]+\epsilon_{WCA}, & r \leq 2^{1/6}r_0 \\ 
0, & r>2^{1/6}r_0 \end{matrix} \right .
\label{WCA}
\end{equation}
We set $\epsilon_{WCA}=1$. In addition to the conservative forces, half of the particles are driven by an external force acting on the center of mass of the particle whose direction changes with a period $\tau$ according to:
\begin{align}
&{\bf F}_{ d}=A\sin\theta \hat e_x+A\cos\theta\hat e_y \label{eq:FexEq} \\
&\theta=2\pi t/\tau. \label{eq:theta}
\end{align}
We characterize the driving forces in terms of the P\'eclet number ($Pe$,) a dimensionless measure of the ratio of advective to diffusive velocity in the system that we define here as $Pe = \frac{A/\gamma}{D_0/r_0}$~\cite{delJunco2018}. For a driven particle, the effect of ${\bf F}_d$ is to cause the particle to orbit in a circle of radius $D_0\beta Pe\tau/(2\pi)$.   For the other half of the particles, ${\bf F}_{d}=0$. 

The second model consists of a mixture of driven LJ particles and undriven WCA particles. We refer to it as the LJ model. The WCA particles move according to Eq.\,\ref{eq:xEOM} and \ref{WCA} with ${\bf F}_{ d}=0$.  The LJ particles move according to Eq.\,\ref{eq:xEOM} with ${\bf F}_{ c,i}$ due to the truncated LJ potential~\cite{Jones1924}:

\begin{equation} 
u(r_{ij})=\left\{ \begin{matrix} 4\epsilon_{LJ} \left[ \left( \frac{r_0}{r_{ij}} \right)^{12}-\left( \frac{r_0}{r_{ij}}\right)^{6}\right] , & r \leq 2.5r_0 \\ 
0, & r>2.5r_0 \end{matrix} \right.
\label{LJ}
\end{equation}
with $\epsilon_{LJ} = 2.25$, and with ${\bf F}_{ d}$ given by Eqs.\,\ref{eq:FexEq} and \ref{eq:theta}. 

Molecular dynamics simulations of both models were performed using a custom Brownian dynamics integrator in LAMMPS~\cite{Plimpton1995}. Results reported here are for square simulation boxes with sides of length $L = 100r_0$ unless otherwise indicated and periodic boundary conditions.  We initiated the simulations by placing a slab $50r_0$ wide of driven particles in the middle of the box, spanning the system in the $y$-direction, so that there were two interfaces of length $L$ along the $y$-direction. 

\begin{figure}
\center
\includegraphics[width=\linewidth]{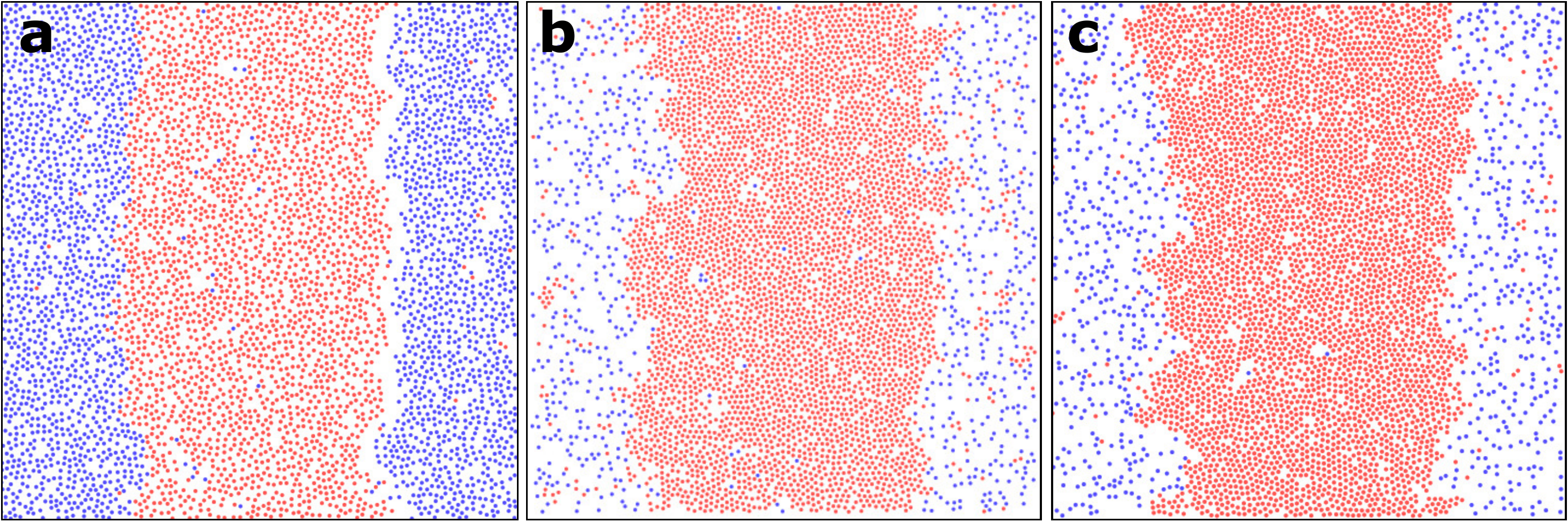}
\caption{{\bf Two models of driven liquids exhibit stable, system-spanning interfaces.} Snapshots of (a) the WCA model with $Pe = 100$, of the LJ model (b) at equilibrium, and (c) with $Pe = 40$ show the slab geometry used in our simulations. Active particles are colored red, and passive particles are colored blue. In the driven cases, a gap is visible at the right interface, which is particularly noticeable in the WCA system. The gap switches from one interface to the other with a period $\tau$.}
\label{fig:snapshots}
\end{figure}

We characterized the phase diagram of the WCA model at a number density $\rho=N/L^2=0.5$ and chose the parameters of the driving force accordingly. At $\tau = 0.1$, the system phase separates in to regions of driven and undriven particles when $Pe \approx 50$ and becomes mixed again at large values of $Pe > 150$, so we chose to simulate interfaces at $\tau = 0.1, \rho = 0.5$, and $Pe = 60, 80, 100$ and 120. In Fig.~\ref{fig:snapshots} we show a snapshot of the system with $Pe = 100$ in the steady state.

For the LJ model we chose the initial density of the slab of LJ particles, $\rho_{LJ} = 0.85$, such that they would exhibit liquid-vapor coexistence in the absence of driving forces~\cite{Smit1991}, and we chose a density of passive WCA particles so that the total density of the system was $0.5$. At equilibrium this results in a liquid phase of LJ particles with a density of $\sim0.72$ in coexistence with a gas of LJ and WCA particles (Fig.~\ref{fig:snapshots}).  We fixed $\tau = 0.1$ and varied $Pe$ from 0 to 80. Examples of steady-state configurations of both models are shown in Fig.~\ref{fig:snapshots}.

The expected relaxation time of the longest-wavelength interface mode was approximated as $\tau_r = L^2 / D$, where $L$ is the length of the interface and $D$ is the diffusion constant of the WCA model in the absence of driving~\cite{delJunco2018}. In the WCA system, we ran each simulation for 10$\tau_r$, discarded the first $\tau_r$ of the trajectory and performed the CWT analysis on the remaining 9$\tau_r$.  In the LJ system, we ran each simulation for 20$\tau_r$, discarded the first 10$\tau_r$ of the trajectory and performed the CWT analysis on the remaining 10$\tau_r$.The center of mass was adjusted in the simulation at intervals of $t_0$ to compensate for drift, and as an extra precaution we also subtracted any center-of-mass motion before analyzing the trajectories.

The box dimension perpendicular to the interfaces, $L_x$, was wide enough that the interfaces were stable along the $y$-direction, and that the width of the interfaces was unrestricted.

\subsection{Interface Current, Density, and Work}

Driven liquids with rotational dynamics can exhibit currents along boundaries and interfaces\cite{Nguyen2014b, VanZuiden2016}. Because of the slab geometry of the present system, any currents would have to be in the $y$-direction. To calculate the particle current, the simulation box was divided in to slices of width $r_0$.  For all particles in a given slice of the box at time $t + t_0$, the displacement $\Delta y = y(t + t_0) - y(t)$ was calculated.  Although there is no velocity in Brownian equations of motion, we report $v_y = \Delta y /t_0$ as an analog of the velocity. The average $v_y$ as a function of $x$ was then calculated by averaging over all of the particles in the slice between $x$ and $x+r_0$ over an interval $\tau_r = 20000t_0$ after the system has reached a steady state. The average $v_y$ in the bulk phase of driven particles was subtracted. 

The density profile of driven particles was measured by dividing in to slices of width $r_0$ and calculating $\rho(x) = N/(L\times r_0)$ in each slice of the box at intervals of $t_0$, where $N$ is the number of particles located in the slice between $x$ and $x+r_0$. The average density profile was obtained by averaging $\rho(x)$ over an interval $\tau_r = 20000t_0$ in the steady state.

We define the work done on the system by the driving forces as~\cite{delJunco2018}
\begin{equation}
\label{eq:work}
\langle \dot{w} \rangle=-\sum_{i=1}^{N} \frac{1}{\tau}\int_0^\tau \dfrac{\langle {\bf F}_{{ c,i}}(t)\rangle\cdot {\bf F}_{{ d,i}}(t)}{\gamma}dt
\end{equation}
where ${\bf F}_{{ c}}$ and ${\bf F}_{{ d}}$ are defined in Eqs.~\ref{eq:xEOM}-\ref{LJ}. This definition of work quantifies the energy input to the system as the driving forces push particles in to one another at each timestep; this energy is subsequently dissipated to the bath as heat. To measure the work in simulations, at each timestep we summed ${\bf F}_{c, i} \cdot {\bf F}_d \Delta t /\gamma$ over all driven particles. This quantity was summed over intervals of $\tau$ and divided by $\tau$ to get $\dot w$. The averaged $\langle \dot w\rangle$ and errors shown in Fig.~\ref{fig:work} are the average and standard deviation of $\dot w$ over 300 periods of $\tau$ after the system has reached a steady state.

\subsection{Capillary Wave Theory and Analysis}

Our analysis of interfacial fluctuations is motivated by CWT~\cite{Bedeaux1985}.  For a flat interface of length $L$, CWT posits that fluctuations in the height of the interface are described by the effective Hamiltonian:
\begin{equation}
H  = \frac{ \sigma }{ 2 } \int_L dx \left| \frac{ dh }{ dx } \right| ^2 
\label{eq:CWTHamiltonian}
\end{equation}
where $h(x)$ is the height of a 1D interface. Using Parseval's identity to take the Fourier transform yields a quadratic Hamiltonian in Fourier space, so we can apply equipartition theorem and obtain an expression for the average height fluctuations of the interface~\cite{Bedeaux1985}:
\begin{equation}
\langle |h(k)|^2\rangle = \frac{k_BT}{L\sigma k^2}.
\label{eq:CWT}
\end{equation}
Here $k$ is a scalar since we are considering straight, 1D interfaces in this work, but Eqs.~\ref{eq:CWTHamiltonian} and \ref{eq:CWT} are easily generalized to higher dimensions~\cite{Bedeaux1985}.  In equilibrium, the $\sigma$ appearing in Eq.~\ref{eq:CWT} should match the surface tension obtained by any other means~\cite{MTOC}. Out of equilibrium that may or may not be the case~\cite{Bialke2015, Patch2018} - nonetheless, if we find that the height fluctuations of the interface scale as $1/k^2$, we can use Eq.~\ref{eq:CWT} to extract $\sigma$ which we may call an effective surface tension~\cite{Bialke2015, Paliwal2017, Derks2006}. We note that in systems where capillary scaling is not obeyed, deviations from $1/k^2$ scaling have been connected to the violation of fluctuation-dissipation theorem - in other words, height fluctuations can still provide insight in to how energy input affects correlations in the system~\cite{Zia1991b}. 

To clearly define the location of the interface, we performed a coarse-graining of snapshots of the system at intervals of $t_0$ by dividing the simulation box up in to a grid with cells $2r_0\times 2r_0$ in dimension, yielding a lattice of dimensions $n \times n$ with $n = L/2$. We assigned a value of 1 to a grid site if it contained at least one driven particle, and a value of 0 otherwise. For the subsequent analysis we only considered one of the two interfaces. We used an image processing algorithm on each frame to extract two contiguous clusters of grid sites, one with value 1 and the other with value 0, separated by an interface. The interface height at $j = y/2$ is the number of sites with value 1 in column $j$.  To obtain $|h(k)|^2$, we took the discrete Fourier transform of $h'(j) = h(j) - \langle h(j) \rangle$. We averaged over all of the snapshots in an interval $\tau_r$ to obtain $\langle |h(k)|^2\rangle$, and checked that the statistics did not change systematically between segments of $\tau_r$. We took the segments to be statistically independent, and we averaged over them to get a second average $\langle \langle |h(k)|^2\rangle\rangle$ - this double average is the value reported in Figs.~\ref{fig:hk_WCA} and ~\ref{fig:LJ}. The error was estimated as the standard deviation of  $\langle |h(k)|^2\rangle$ between the analyzed segments. The code used for the analysis is available upon request.

In the WCA system where the scaling of $\langle |h(k)|^2\rangle$ was not $1/k^2$, we extracted the scaling exponent by fitting the linear part of a log-log plot of $\langle |h(k)|^2\rangle L$ vs $1/k^2$, judged by eye from the data in Fig.~\ref{fig:hk_WCA}. Where applicable, the surface tension was extracted by fitting $\langle |h(k)|^2\rangle$ according to Eq.~\ref{eq:CWT} over a range from $k_{min}$ to $k_{max}$, where $k_{max}$ was defined as the largest value of $k$ for which $\langle |h(k)|^2\rangle$ was greater than the coarse-graining length of the system and $k_{min}$ was defined as the the smallest value of $k$ for which $1/k^2$ was a good fit to $\langle |h(k)|^2\rangle$, judged by eye from the data in Fig.~\ref{fig:LJ}.

\section{Results}

\subsection{Phase Separation}

First we briefly recapitulate the mechanism of phase separation in the WCA model, which was explored in more detail in Refs.~\citenum{delJunco2018} and \citenum{Han2017}. In Ref.~\citenum{delJunco2018}, we found that the driving forces do work (as defined in Eq.~\ref{eq:work}) on the system by inducing collisions between particles. These collisions result in an increased diffusion coefficient which scales roughly proportional to $w$, the amount of work done per period of driving, which in turn scales as $Pe^2$. 
Because the driven particles are always in phase, in a region with only driven particles or only undriven particles, the nonequilibrium forces do not induce any collisions. The work done and therefore the diffusion coefficient thus depend on the local composition, and particles diffuse faster out of regions with mixed configurations than back in to them. If the gradient of the diffusion with respect to composition is sufficiently high, this results in phase separation of driven and undriven particles. 

This mechanism is similar to what has been proposed for systems that undergo laning (separation of two types of particles moving in opposite directions in to lanes parallel to their velocity vectors)~\cite{Klymko2016} and stripe formation (separation in to stripes perpendicular to the direction of periodic forcing)~\cite{Mullin2000, Pooley2004, Wysocki2009}. In both cases, the differential mobility of the particles in the presence of the other particle type leads to separation. 
We note that this mechanism of phase separation depends on a high degree of synchronization between the displacement vectors of the driven particles of each type - in our case, the driving force on all driven particles is the same (Eq.~\ref{eq:FexEq}), so that all the driven particles are in phase. If the directions of the driven particles are not correlated, for instance if we assigned random phases to each driven particle, phase separation of the kind seen here would not occur. Instead, for sufficiently high $Pe$ and slowly changing particle direction, we would expect motility-induced phase separation~\cite{Liebchen2017, Cates2015}.  

Similar to Refs.~\citenum{Mullin2000, Pooley2004, Wysocki2009, Vissers2011a}, in the WCA system there is a gap at one of the interfaces between the red and blue particles (Fig.~\ref{fig:snapshots}). The location of the gap switches periodically from one interface to the other. This is because the red (driven) particles effectively occupy a larger volume than the blue particles and push them out of the way when the driving force pushes red and blue particles in to one another. When the force changes directions, the red particles move en masse away from the blue particles, but diffusion is not fast enough for the blue particles to fill the space left by the retreating red particles, so a gap opens up.  

\subsection{Currents Along the Interface}

\begin{figure}
\includegraphics[width=\linewidth]{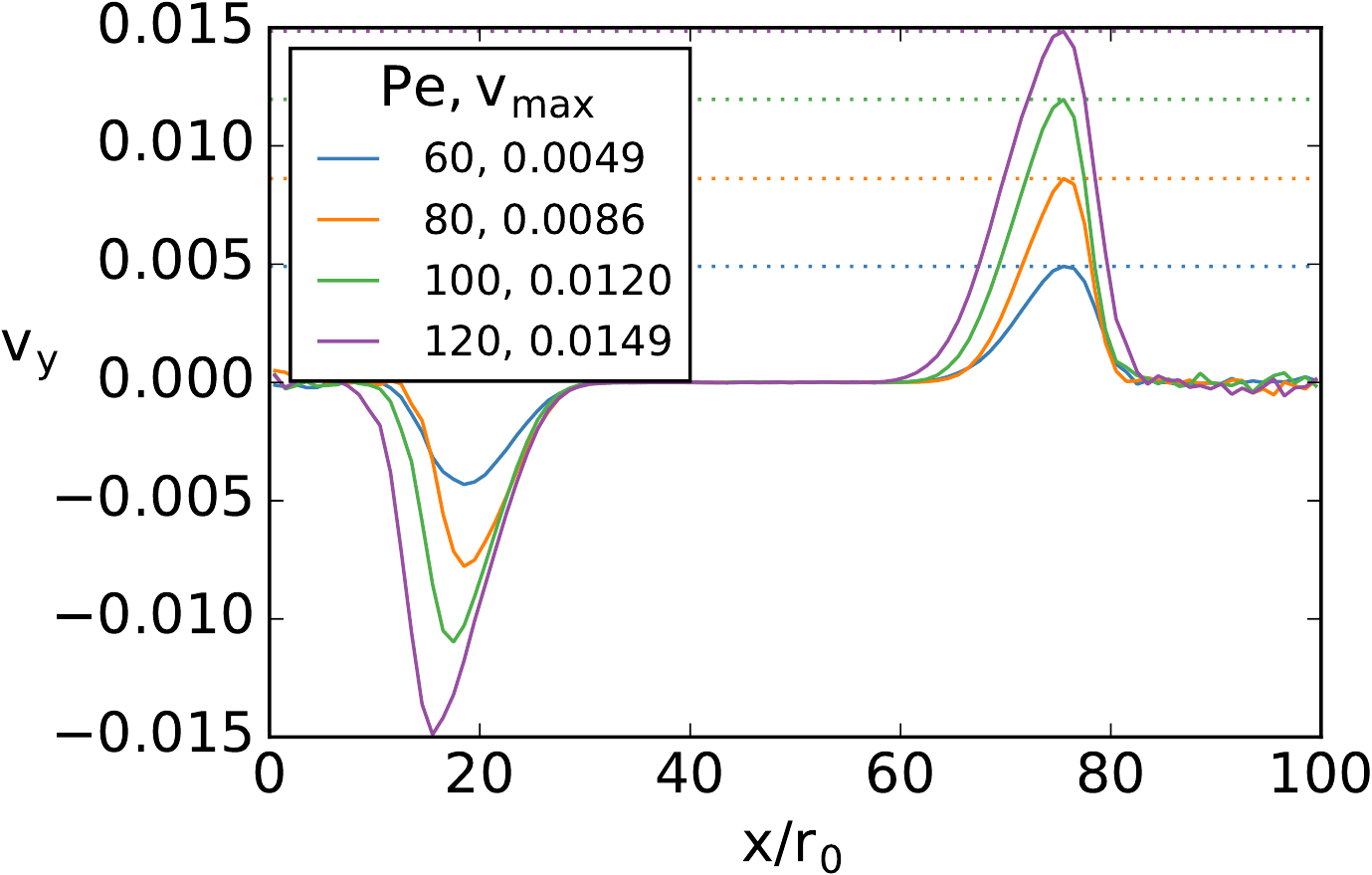}
\caption{{\bf There is a net particle current along the interface in the driven WCA system.} $v_y$, defined in Methods, quantifies the current along the interface in the $y$-direction. The maximum value of $v_y$ scales roughly linearly with $Pe$. Since the average position of the interface varies between simulations, the curves have been shifted in the $x$-direction to facilitate comparison.} 
\label{fig:velocityprof}
\end{figure}

Measuring the $y$-direction displacement of particles in the WCA model reveals that there are particle currents along the interface.  In Fig.~\ref{fig:velocityprof} we show that the direction of the flow is chiral - by which we mean that it moves in only one direction along the interface as determined by the direction of orbit of the driven particles - and that its maximum value is roughly linear in $Pe$. This feature distinguishes interfaces in this system from ones previously studied in MIPS-type systems with WCA~\cite{Bialke2015, Lee2017, Patch2018} or LJ~\cite{Paliwal2017} interactions, where no flows exist in the steady state due to the random orientation of the active forces. 

\subsection{Scaling of Interface Fluctuations and Surface Tension}

\begin{figure}
\includegraphics[width=\linewidth]{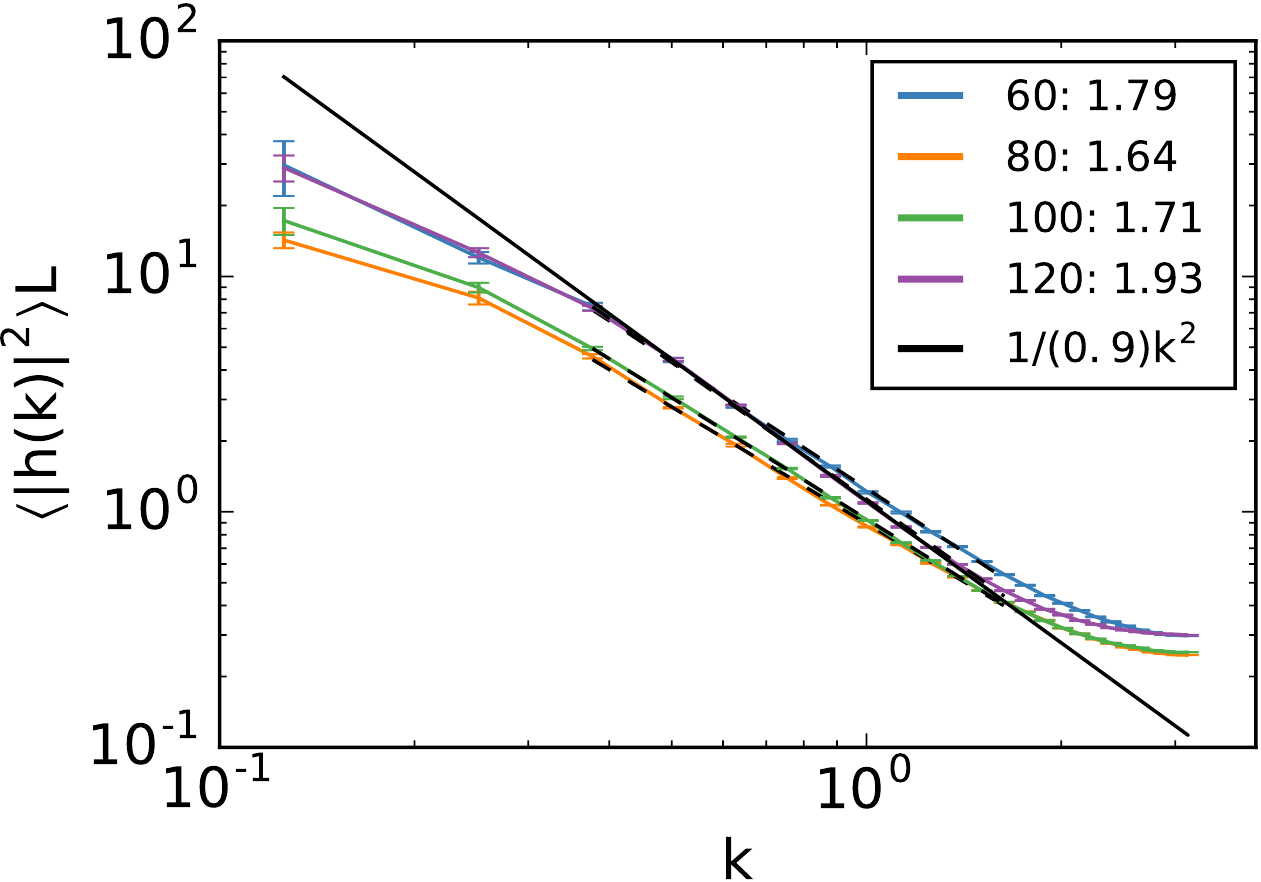}
\caption{{\bf The WCA model exhibits capillary scaling for ${\mathbf{Pe = 120}}$.} Scaling of interface modes ($\langle |h(k)|^2\rangle$) multiplied by interface length ($L$) in the WCA model as a function of $k$, for interfaces of length $100r_0$ (solid colored lines). The legend indicates values of $Pe$ and of the scaling exponent $\alpha$, as in $\langle |h(k)|^2\rangle \propto k^{-\alpha}$, obtained by fitting over the region indicated by dashed black lines, in the format ($Pe: \alpha$). The fluctuations for $Pe = 60, 120$ are larger than for $Pe = 80, 100$. In this system, $Pe = 60$ is close to the point where the system first phase separates, while $Pe = 120$ is close to the point where the system becomes mixed again. $Pe = 80$ and 100 are further inside the bulk of the phase separated region of the phase diagram. For $Pe = 120$, the scaling of fluctuations is close to the $1/k^2$ signature of capillary wave theory over roughly an order of magnitude in $k$, so for this case we also fit a line (solid black line) $\propto 1/k^2$ to calculate an effective surface tension $\sigma/k_BT = 0.9$. The error bars are negligibly small except for at $k < 0.3$.}
\label{fig:hk_WCA}
\end{figure}

Based on studies of driven lattice gases~\cite{Leung1988b, Leung1993} we might expect currents parallel to the interface in the WCA model to cause deviations of the interface height fluctuations from capillary scaling. Indeed, most of the parameters that we studied do not exhibit capillary scaling. However, for $Pe = 120$, the spectrum of interface fluctuations has an exponent close to -2, but only over roughly an order of magnitude of wavenumbers (Fig.~\ref{fig:hk_WCA}). For this value of $Pe$ we calculated an effective surface tension of $\sigma/k_BT = 0.9$. For lower values of $Pe$, $\langle |h(k)|^2\rangle$ decreases less rapidly with $k$ than predicted by CWT. The scaling exponents extracted from fits are shown in Fig.~\ref{fig:hk_WCA}; however we emphasize that these should not be interpreted as analytical exponents resulting from some underlying physics. We note that the system undergoes a reentrant mixing transition as the value of $Pe$ is increased~\cite{delJunco2018}. In particular, the point $Pe=120$ is close to the rentrant transition. Due to this, we were unable to systematically probe the effects of increasing the driving force amplitude on the interfacial fluctuations. 

At all values of $Pe$, fluctuations for the smallest ($k < 0.4$ in Fig.~\ref{fig:hk_WCA}) and largest ($k > 2$) wavevectors do not follow the same trend as the rest of the data. At large wavevectors $\langle |h(k)|^2\rangle$ flattens out as a result of the lower limit on fluctuations set by our coarse-graining of the system. To test whether the flattening at small wavevectors was a real feature or an artefact of the finite simulation time, we simulated a trajectory with $Pe = 120$ and $L = 200r_0$ for 8 times longer than the $L = 100r_0$ simulations. There, $1/k^2$ scaling persists to larger wavelengths, suggesting that the fall-off is indeed due to the simulation time.

The results in Fig.~\ref{fig:hk_WCA}, as well as previous results on interfaces in active systems~\cite{Paliwal2017}, suggest that driving can change the effective surface tension and modify the statistics of interfaces in nonequilibrium liquids.  Studying these effects in a systematic way is complicated in the WCA model by the fact that the driving has a non-monotonic effect on the interface statistics over a relatively narrow range of values of $Pe$, but a linear effect on the magnitude of the particle flow along the interface. In addition, since this system cannot phase separate in the absence of driving, there is no reference equilibrium interface to compare the driven interfaces to. To address this issue, we use the LJ model, which exhibits liquid-vapor coexistence at equilibrium. Interfaces in LJ liquids have been well-studied and are known to exhibit capillary scaling~\cite{Sides1999, MTOC}, so the LJ model provides a clear reference point that is lacking in the WCA model, and moreover, we can study the effect of driving forces starting well below $Pe = 60$. 

\begin{figure}
\centering
\includegraphics[width = \linewidth]{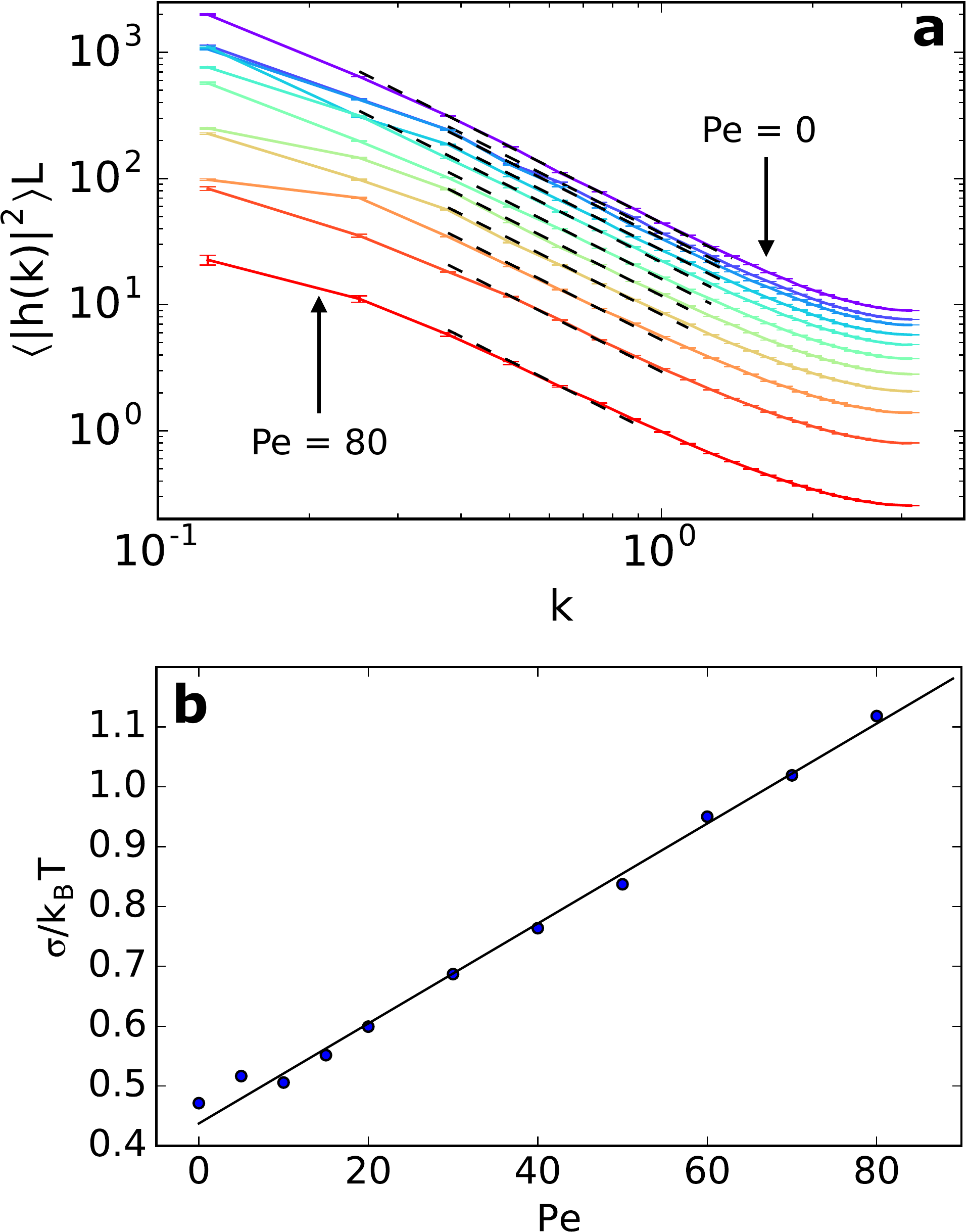}
\caption{{\bf Driving increases the surface tension linearly and modifies the scaling of interface fluctuations of LJ particles.} (a) Scaling of interface modes ($\langle |h(k)|^2\rangle$) multiplied by interface length $L$ in the LJ model as a function of $k$, for interfaces of length $100r_0$. The curves for $Pe = 0 - 70$ have been offset to make it easier to see that the range of $k$ for which $\langle |h(k)|^2\rangle$ scales as $1/k^2$ is largest close to equilibrium and becomes smaller as $Pe$ increases; the $Pe = 80$ curve is not offset to show that the magnitude of fluctuations is comparable to the WCA system. Black lines are $\propto 1/k^2$ and show the range of the fits used to extract $\sigma$; in this range, the error bars are very small. (b) Surface tension (measured from the fits of $\langle |h(k)|^2\rangle$) as a function of $Pe$, with a fit showing the linear correlation between $\sigma$ and $Pe$.}
\label{fig:LJ}
\end{figure}

We first verified that the LJ model produced the expected behavior at equilibrium. We show in Fig.~\ref{fig:LJ} that at $Pe = 0$, the LJ system exhibits capillary fluctuations with a value of the surface tension that is in reasonable agreement with literature values~\cite{Santra2009}. The range of capillary scaling in $k$ is again limited from above by the coarse-graining length and from below by the simulation time.  We then measured the effect of driving the system with $Pe$ ranging from  $5 - 80$.  We find that the surface tension increases linearly over the whole range of $Pe$ (Fig.~\ref{fig:LJ}). Based on our own previous work~\cite{delJunco2018}, which shows that driving can stabilize interfaces in this system, and on other studies of surface tension in driven systems~\cite{Paliwal2017}, we expected an increase in surface tension. However, those results do not indicate that the increase would be linear and persist over the entire range of $Pe$ investigated here, which is an order of magnitude larger than in Ref.~\citenum{Paliwal2017}.  In the following section, we present phenomenological arguments and simulation data that partially account for this observation.

\subsection{Origin of the Increase in Surface Tension}

Despite our heuristic understanding (summarized in the first part of the Results) of how the driving forces in our model cause phase separation and therefore how they can create interfaces (in the WCA model) or stabilize them (in the LJ model) by increasing the surface tension, it is not clear why the increase should be linear in $Pe$. Surface tension arises due to an imbalance in the forces on particles near to the interface.  We now consider two ways that time-dependent driving forces of the kind studied here can magnify this force imbalance, and whether these can explain the observed doubling of the surface tension (Fig.~\ref{fig:LJ}).

\begin{figure}
\includegraphics[width = 0.6\linewidth]{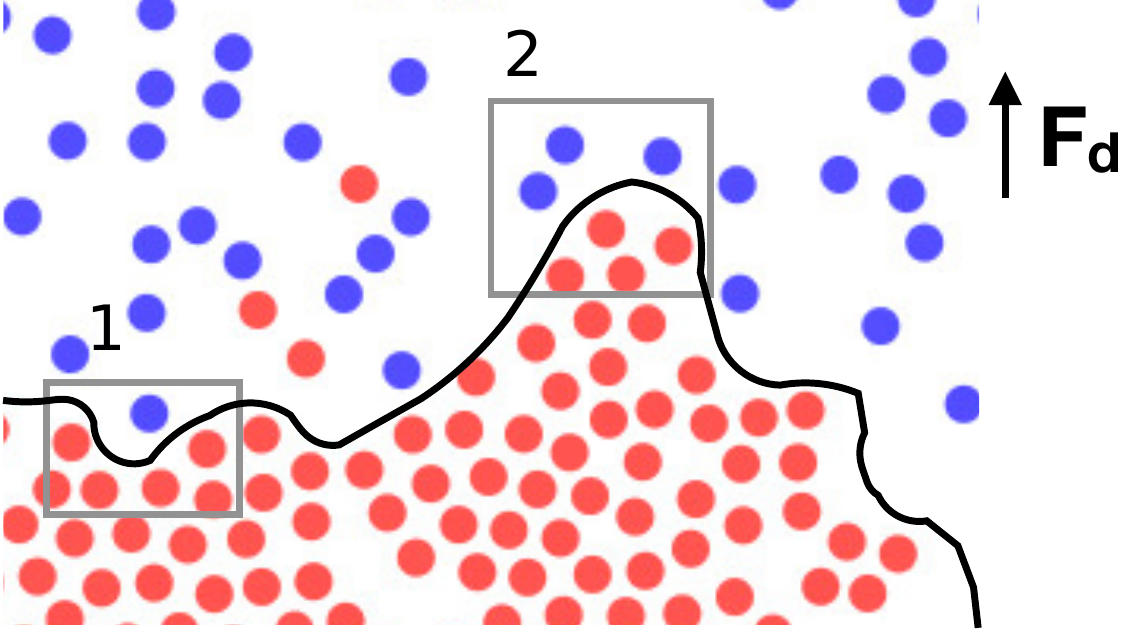}
\caption{{\bf WCA particles exert a force proportional to $\mathbf{Pe\nabla^2 h}$ on LJ particles near the interface.} At the moment of the snapshot, all red LJ particles are pushing up on the blue WCA particles with a force ${\bf F}_d\propto Pe$. In the box labeled 1, where the curvature is positive, LJ particles experience an opposing conservative force from 1 WCA particle. In the box labeled 2, where the curvature is positive, LJ particles experience an opposing conservative force from 3 WCA particles. On average, this leads to a force on the interface $\propto Pe\nabla^2 h$.}
\label{fig:schematic}
\end{figure}

First, we propose that the driving forces can cause the undriven WCA particles to exert a restoring force on regions of the interface with high curvature.  To see why, consider a section of the interface like the one shown in Fig.~\ref{fig:schematic}. All LJ particles at the interface experience a force ${\mathbf F}_d\propto Pe$ that pushes them in to undriven WCA particles. In the linear response regime, WCA particles will push back with a conservative force also proportional to $Pe$~\cite{delJunco2018}. A driven particle at the interface will therefore feel a downward force proportional to $Pe$ and to the number of undriven particles in its neighborhood~\footnote{Although the force exerted by the driven particle is not always pointed straight in to the undriven phase as illustrated in Fig.~\ref{fig:schematic} - it rotates according to Eq.~\ref{eq:FexEq} - when the driven particles are moving away from the undriven particles they exert no force on them, since the driven-undriven particle interactions are purely repulsive. This results in the gaps that we observe at the interface in the WCA system. To a first approximation, we therefore assume that the most important contribution to ${\mathbf F}_d$ points out normal to the interface and restoring force ${\mathbf F}_d$ points back down.}.
As we illustrate in Fig.~\ref{fig:schematic}, if the driven particle is at a point with negative curvature, it is surrounded by more undriven particles than if it is at a point with positive curvature. Thus, the excess downward force on the interface is proportional to the curvature: $\langle F_c\rangle_{int}\propto Pe\nabla^2 h$. Combining this argument with the CWT Hamiltonian in Eq.~\ref{eq:CWTHamiltonian} we can write down a phenomenological equation of motion for $h(x)$:
\begin{equation}
\frac{\delta h}{\delta t} = \frac{\sigma}{2} \nabla^2 h(x) + Pe \nabla^2 h(x) + \eta(x, t),
\label{eq:CWTEOM}
\end{equation}
where $\eta$ is a white noise with statistics $\left<\eta(x, t)\right>=0$ and $\left<\eta(x, t)\eta(x',t')\right> = 2k_BT\delta (x - x') \delta(t-t')$. We immediately see that this will result in an apparent surface tension $\propto Pe$.  

\begin{figure}
\includegraphics[width = \linewidth]{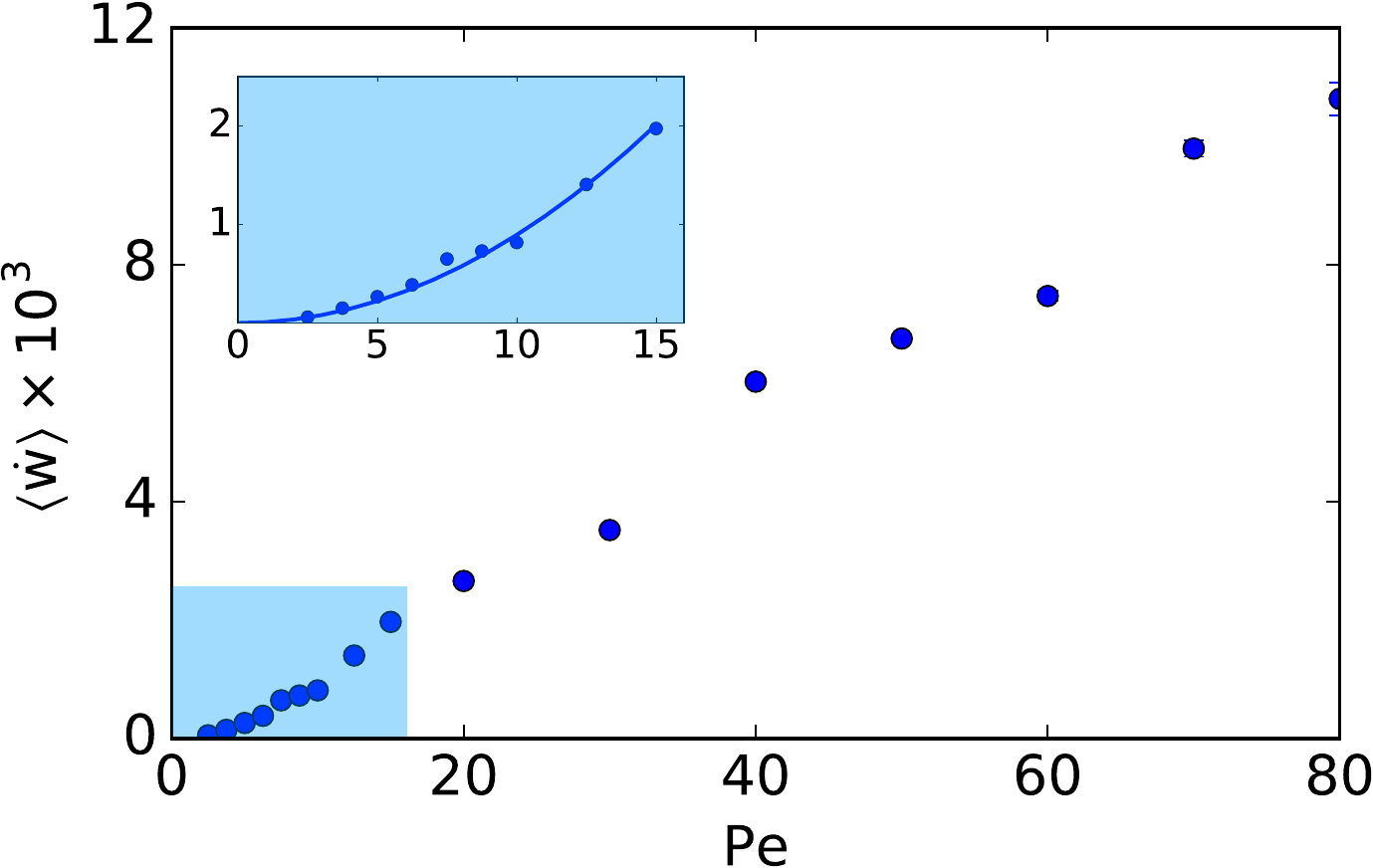}
\caption{{\bf The rate of work done on the system by driving forces scales linearly with $Pe$ for $Pe > 15$.} In the inset we show that for values of $Pe \leq 15$ the work scales as $Pe^2$, in agreement with the results of Ref.~\citenum{delJunco2018}. Error bars are smaller than the points except for at $Pe = 80$.}
\label{fig:work}
\end{figure}

For this picture to correctly explain our observations, $\langle {\bf F}_c\rangle$ must scale with $Pe$, which implies that the work done on the system at the interface by the driving forces should scale as $Pe^2$, since the work is proportional to ${\bf F}_c\cdot {\bf F}_d$ (Eq.~\ref{eq:work}). Motivated by earlier work on this system in which we found that work in a region of mixed driven and undriven particles scales as $Pe^2$, we hypothesized that this could also be the case at the interface. To check whether this is indeed the case, we measured the work in the system according to Eq.~\ref{eq:work}. Work can only be done where there are driven and undriven particles in contact, so although we measured the work in the whole system, the small number of LJ particles in the WCA bulk and vice-versa (Fig.~\ref{fig:snapshots}) ensures that the interfacial region provides the important contribution to the total work. Contrary to our hypothesis, we show in Fig.~\ref{fig:work} that in the LJ system the work is only quadratic in $Pe$ for $Pe < 15$, and then follows a linear trend up to $Pe = 80$. 
This means that $\langle F_c\rangle_{int}\propto Pe\nabla^2 h$ can only partially explain the linear scaling of $\sigma$ with $Pe$.

\begin{figure}
\includegraphics[width = \linewidth]{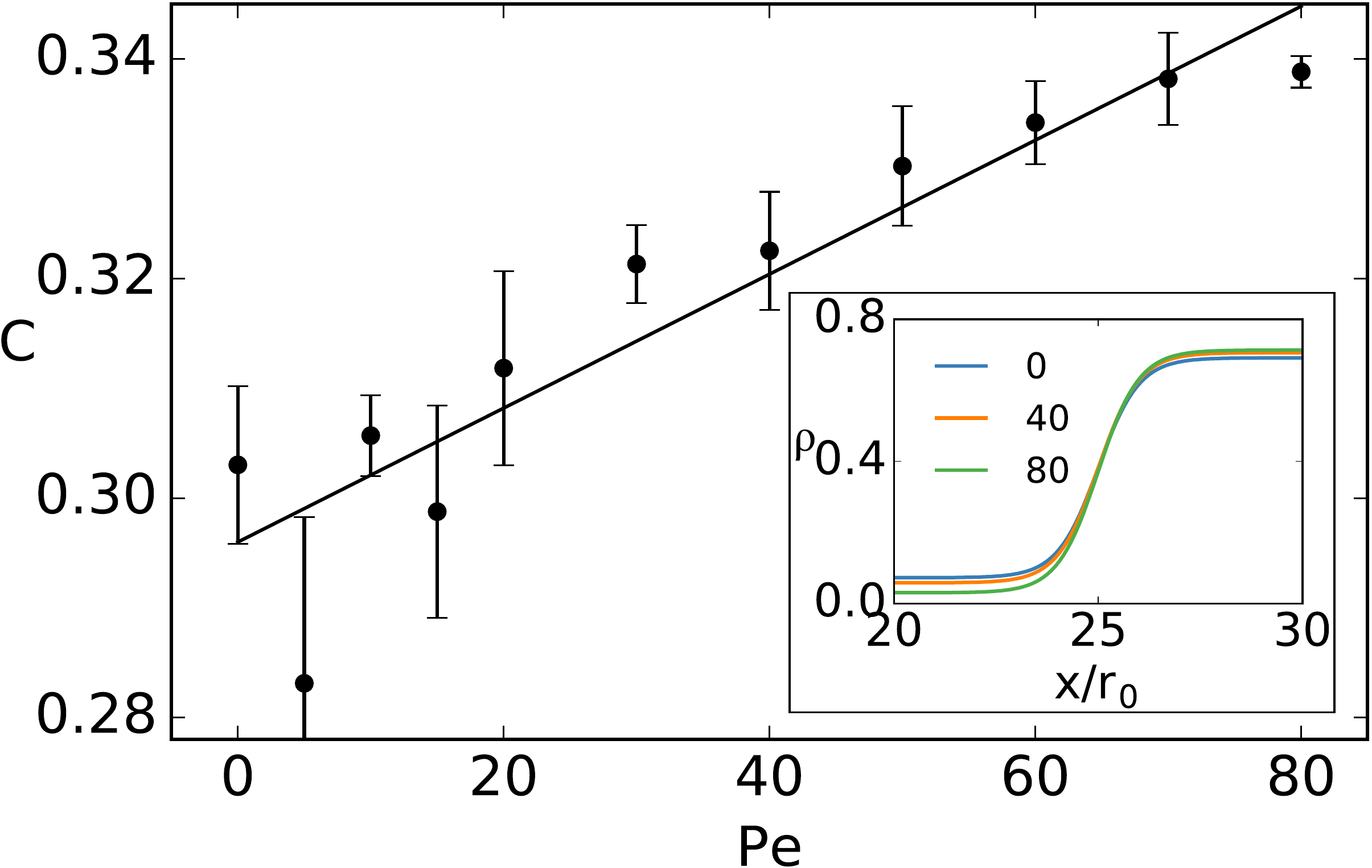}
\caption{{\bf The density gradient near the interface scales linearly with $\mathbf{Pe}$.} The slope of the density near the interface, given by $C$ as defined in Eq.~\ref{eq:tanh}, as a function of $Pe$. Error bars are the standard deviation of the values of $C$ obtained from fitting density profiles of four independent segments of the simulation of length $\tau_r$.  (Inset) An example of the fits of the density of LJ particles to Eq.~\ref{eq:tanh} for $Pe = 0, 40, 80$ shows the liquid density increasing and the gas density decreasing with increasing $Pe$.}
\label{fig:density}
\end{figure}

Another way that the driving forces can modify the force imbalance is by increasing the density of the LJ liquid phase, so that the imbalance in attractive forces is magnified.  We measured the density of the driven LJ particles as a function of position to see if there was a significant change. Indeed, as $Pe$ is increased the density of LJ particles in the liquid phase increases, and the density of LJ particles in the gas phase decreases. To quantify the change we fit the density of the left interface to a hyperbolic tangent function of the form:
\begin{equation}
\rho(x) = C\tanh(x - x_0) + b.
\label{eq:tanh}
\end{equation}
where $C, b$ and $x_0$ are fitting parameters. Assuming this form for the density, the force imbalance on a particle located at the interface is proportional to $C$, so $C$ should predict the increase in surface tension due to the change in density. In Fig.~\ref{fig:density}, we show that $C$ increases roughly linearly with $Pe$. However, the change in $C$ is only on the order of 15\% and cannot explain the full increase in the surface tension that we observed.  The driving forces must therefore have effects on the interface in addition to a force proportional to $Pe\nabla^2 h$ and an increase in density; what these effects might be remains an open question. Importantly, in both of these arguments we ignored the time dependence of the driving forces. The time dependence is what causes currents at the interface, which are expected to affect fluctuations~\cite{Leung1993}. We therefore expect that it will be necessary to take in to account time-dependent affects such as coupling between interface modes to account for our results.

\section{Conclusions}

In this simulation study we presented results regarding the surface tension and statistics of interfacial fluctuations in two closely related systems of driven particles: one where all particles have repulsive WCA interactions and half are driven, and a second where the driven particles have attractive LJ interactions. The WCA system is phase separated for a range of P\'eclet numbers from approximately $Pe = 50-150$.  Over this range the interfaces exhibit chiral particle currents parallel to the interface whose velocity is proportional to $Pe$. At one value of $Pe$ near to the reentrant phase transition, height fluctuations of the interface exhibit the $1/k^2$ scaling that is a signature of capillary wave theory. For other values of $Pe$, the spectra of height fluctuations are inversely proportional to $k$ but less steep than $1/k^2$. In the system with LJ interactions, stable interfaces with capillary scaling already exist at equilibrium. Upon driving, we found that capillary scaling persists and that surface tension increases linearly over two orders of magnitude in $Pe$  - from small values in the linear response regime to well above the value of $Pe$ required for phase separation in the WCA system. 

The driving force in our system can be reproduced in an experiment using rotating magnetic fields~\cite{Han2017}. Our findings therefore suggest a way of controlling the surface tension of assemblies of particles from a distance, without the need to change any properties of the particles. However, although we discussed two possible explanations for the excess force imbalance at the interface that causes the increase in the surface tension with $Pe$ (a force proportional to the curvature of the interface induced by the driving forces, and the increased density gradient of the LJ particles), neither captures the doubling in the magnitude of the surface tension that we observed.  Our work thus poses the challenge of fully explaining how the system channels the energy input at the smallest possible length scale in to modes at the interface that span the largest length scale of the system, which we expect will require a theory taking in to account the genuinely non-equilibrium nature of the steady state. This understanding will be necessary to fully control the surface tension of experimental particle systems.

\section{Acknowledgements}

Thanks to Glen Hocky and Bodhi Vani for helpful comments on this draft. This work was partially supported by the University of Chicago Materials Research Science and Engineering Center, which is funded by the National Science Foundation under award number DMR-1420709. CdJ and SV also acknowledge support from the Sloan Fellowship and the University of Chicago.

%\bibliography{/Users/claradeljuncooffice/Documents/library.bib}

%merlin.mbs apsrev4-1.bst 2010-07-25 4.21a (PWD, AO, DPC) hacked
%Control: key (0)
%Control: author (8) initials jnrlst
%Control: editor formatted (1) identically to author
%Control: production of article title (-1) disabled
%Control: page (0) single
%Control: year (1) truncated
%Control: production of eprint (0) enabled
%

\end{document}